# Photonic Circuit of Arbitrary Non-Unitary Systems

Hussein Talib[1*], Phillip D. Sewell[1], Ana Vukovic[1], and Sendy Phang[1*]

[1]George Green Institute for Electromagnetics Research, Faculty of Engineering, University of Nottingham, Nottingham, NG9 1EY, UK
*hussein.talib@nottingham.ac.uk, Sendy.Phang@nottingham.ac.uk

**Abstract**

A design framework to implement non-unitary input-output operations to a practical unitary photonic integrated circuit is described. This is achieved by utilising the cosine-sine decomposition to recover the unitarity of the original operation. The recovered unitary operation is decomposed into fundamental unitary building blocks, forming a photonic integrated circuit network based on directional couplers and waveguide phase shifters. The individual building blocks are designed and optimised by three-dimensional full-wave simulations and scaled up using a circuit approach. The paper investigates the scalability and robustness of the design approach. Our study demonstrates that the proposed approach of performing unitary matrix completion can be applied to any arbitrary matrices. This design approach allows for implementation of non-unitary operations to perform various linear functions in neuromorphic photonics for computing, sensing, signal processing and communications.

# 1. Introduction

Photonic integrated circuits are a promising platform for implementing a photonic analogue of electronic signal processing devices and thus exploit the full advantage of speed and capacity that optical signal offers. In the earlier research, metaheuristic approaches using optimisation-evolutionary algorithms have been successfully employed in developing very compact and high-performance application-specific photonic integrated circuit devices for arbitrary input-output operations (Sewell et al., 2007) (Vukovic et al., 2010) (Piggott et al., 2015) (Phang et al., 2020). However, the photonic circuit design process, using these metaheuristic approaches, is computationally expensive requiring long simulations and high computational demands. Additionally, it results in devices with unstructured "messy" architectures lacking an underlying topology.

Recent developments have shifted towards a more methodical approach to photonic circuit development based on mathematical linear-system decomposition methods (Bogaerts et al., 2020). Using such a method, a photonic circuit performing any *linear input-output unitary operation* can be implemented in a structured topology, such as triangular (Reck et al., 1994), rectangular (Clements et al., 2016), diamond (Shokraneh et al., 2020), and Bokun (Mojaver et al., 2023), and employing simple unit *building block*, such as beam splitters (Miller, 2013), Mach-Zehnder interferometers (Pérez et al., 2020), ring resonators (Yi et al., 2021) (Mosses & Prathap, 2023) and tuneable couplers (Pérez et al., 2018) (Pérez-López et al., 2019). By using structured topologies, it becomes possible to build a single device that can be repurposed depending on the target operation, enhancing flexibility and efficiency in various applications (Bogaerts et al., 2020).

The notable increase in interest in photonic circuits capable of performing linear operations stems from their fundamental functionalities, specifically multiply-and-accumulate (MAC) operations found applications in signal processing, hashing operations, and artificial neural networks (Zhang et al., 2021)(Chang et al., 2018). Photonics-based MAC devices offer significant potential benefits in terms of speed and energy efficiency (Wang et al., 2021). Miller described the principles to realise such a photonic MAC device with a unidirectional, bidirectional, and feedback system where the unitary matrices are implemented using multiport interferometers and the diagonal matrix is implemented through modulators (Miller, 2013). Other shows a methodology to represents a linear operation in terms of multimode spatial beam transmission by engineering the intermodal crosstalk transmission matrix (Dhand & Goyal, 2015) . A design of a photonic waveguide mesh circuit that implemented arbitrary system is considered in (Pérez & Capmany, 2019) (Sanchez et al., 2022) (Chen et al., 2020) (López et al., 2020) where cells consisting of connected waveguides in triangular, square, or hexagonal shape are used to construct the circuit that implements an arbitrary photonic system.

While the linear decomposition methods have enabled the implementation of any unitary input-output operations, emerging applications often require non-unitary input-output operation, for example the linear read-out weight matrix found in photonic reservoir computing (Phang et al., 2020) (Anufriev et al., 2022) (Phang, 2023). To implement such non-unitary operations using unitary building blocks, an intermediate mathematical procedure, such as Singular Value Decomposition (SVD) (Miller, 2013) and matrix completion (Tang et al., 2024), is needed to transmute the target non-unitary operation into a unitary one.

In the current work, we propose an alternative mathematical procedure to embed the target non-unitary operation into a unitary operation using cosine-sine decomposition (CSD). The proposed CSD approach works by expanding the target operation into a higher-dimensional representation to recover its unitarity. Although the proposed CSD approach results in a larger circuit, it preserves the topology of the circuit network and can be implemented in a single device. In contrast, the direct SVD approach (Miller, 2013) produces a more compact circuit but requires a connecting block, which changes the topology and results in a more complex overall circuit. Furthermore, to demonstrate the CSD unitary completion approach, we developed a photonic MAC device using a rectangular architecture with directional couplers equipped with an S-bend waveguides as a building block. The parameters of these blocks are designed following the linear diagonalisation using the successive nulling algorithm (Clements et al., 2016) to map the reconstructed unitary matrix into a realisable circuit. The resulting circuits were validated using a full-wave three-dimensional Finite Element Method (FEM) implemented in COMSOL (COMSOL, 2023). To further scale up the photonic circuit networks, circuit-based modelling was used, utilising the building blocks' scattering matrices computed by the full-wave simulation.

The paper is structured as follows: Section 2 describes the framework used in the current work to implement a photonic circuit using the CSD and successive nulling algorithm approach. Section 3 demonstrates the application of the design framework, for an arbitrary 2-by-2 random matrix and investigates the scalability and robustness of the design framework. Section 4 summarises the presented work.

## 2. Design Framework

This section describes the design framework used in the present work to develop a photonic integrated circuit based on a directional coupler building block. Figure 1 gives a schematic of the design framework, which is comprised of two mathematical procedures, namely cosine-sine decomposition (CSD) and successive nulling algorithm. The CSD is used to transform an arbitrary non-unitary matrix **M** into a higher-dimensional unitary matrix **U**. Recovering the unitarity property is crucial, as it allows for the use of successive nulling algorithm to map the matrix **U** to a set of practical design parameters of a directional coupler namely, phase shift $\phi$ and coupling constant, $\kappa$.

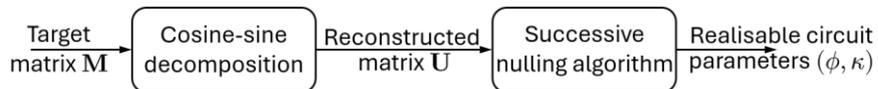

**Figure 1**: Design framework based on cosine-sine decomposition and successive nulling algorithm approach. The design framework yields the required phase $\phi$ and coupling constants $\kappa$ for each of building blocks.

To allow the unitary completion of the target matrix $\mathbf{M} \in \mathbb{R}^{r_M \times c_M}$, with $\{r_M, c_M\} > 1$, one needs to consider the largest singular value of **M** using CSD method. As shown in (Golub & Van Loan, 2013), a sufficient condition to construct a unitary matrix **U** from a given matrix **M** is that the singular values of **M** are upper bounded by 1. Thus, let us consider the matrix $\widetilde{\mathbf{M}} = \mathbf{M}/\hat{\rho}$, which is a real matrix whose largest singular value is one, normalised by $\hat{\rho}$ which is the largest singular value of matrix **M**. Practically, the effect of the common normalisation factor $1/\hat{\rho}$ can be reversed by using an optical amplifier (or attenuator) at the output ports with gain of $\hat{\rho}$.

Recalling the general CSD theorem (Paige & Wei, 1994), which states that for any 2-by-2 partitioning of a unitary matrix

$$\mathbf{U} = \begin{bmatrix} \mathbf{U}_{11} & \mathbf{U}_{12} \\ \mathbf{U}_{21} & \mathbf{U}_{22} \end{bmatrix} \begin{matrix} r_1 \\ r_2 \end{matrix}, \quad \begin{matrix} c_1 & c_2 \end{matrix} \tag{1}$$

there exist unitary matrices $\mathbf{L}_1, \mathbf{L}_2, \mathbf{R}_1, \mathbf{R}_2$ such that,

$$\mathbf{L}^T \mathbf{U} \mathbf{R} = \begin{bmatrix} \mathbf{L}_1^T & \\ & \mathbf{L}_2^T \end{bmatrix} \begin{bmatrix} \mathbf{U}_{11} & \mathbf{U}_{12} \\ \mathbf{U}_{21} & \mathbf{U}_{22} \end{bmatrix} \begin{bmatrix} \mathbf{R}_1 & \\ & \mathbf{R}_2 \end{bmatrix} = \mathbf{D}, \tag{2}$$

where

$$\mathbf{D} \equiv \begin{bmatrix} \mathbf{D}_{11} & \mathbf{D}_{12} \\ \mathbf{D}_{21} & \mathbf{D}_{22} \end{bmatrix} = \left[ \begin{array}{c|c} \begin{matrix} \mathbf{C} \\ \mathbf{0}_C \\ \mathbf{S} \end{matrix} & \begin{matrix} \mathbf{S} \\ \mathbf{1}_S \\ -\mathbf{C} \end{matrix} \\ \mathbf{1}_S & \mathbf{0}_C^T \end{array} \right],$$

$$\mathbf{C} = \text{diag}(\zeta_1, \zeta_2, \cdots), \quad \text{where } 1 \geq \zeta_1 \geq \zeta_2 \geq \cdots \geq 0,$$
$$\mathbf{S} = \text{diag}(\xi_1, \xi_2, \cdots), \quad \text{where } 0 \leq \xi_1 \leq \xi_2 \leq \cdots \leq 1,$$
$$\mathbf{C}^2 + \mathbf{S}^2 = \mathbf{1}. \tag{3}$$

In (1)-(3), $r_i$ and $c_j$ denotes the number of row and columns of the submatrix $\mathbf{U}_{ij}$, respectively, and $\mathbf{0}_C$ and $\mathbf{1}_S$ are matrices of zeros and identity, respectively. Their size depends on the partitioning of unitary matrix **U** as shown in (1). The diag(·) denotes a diagonal matrix. It can be asserted, from (2), that the unitary matrices $\mathbf{L}_1, \mathbf{L}_2, \mathbf{R}_1, \mathbf{R}_2$ are the singular value decomposition of each of the four submatrices **D**,

$$\mathbf{U} = \mathbf{L} \mathbf{D} \mathbf{R}^T = \begin{bmatrix} \mathbf{L}_1 \mathbf{D}_{11} \mathbf{R}_1^T & \mathbf{L}_1 \mathbf{D}_{12} \mathbf{R}_2^T \\ \mathbf{L}_2 \mathbf{D}_{21} \mathbf{R}_1^T & \mathbf{L}_2 \mathbf{D}_{22} \mathbf{R}_2^T \end{bmatrix} \equiv \begin{bmatrix} \mathbf{U}_{11} & \mathbf{U}_{12} \\ \mathbf{U}_{21} & \mathbf{U}_{22} \end{bmatrix} \begin{matrix} r_1 \\ r_2 \end{matrix}. \quad \begin{matrix} c_1 & c_2 \end{matrix} \tag{4}$$

The unitary matrix completion in the present work makes use of the decomposition properties given by the CSD theorem in (2)-(4). This is done by variable substitutions such that

$$\mathbf{U} \leftarrow \begin{bmatrix} \overset{c_M}{\mathbf{M}} & \overset{r_M}{\mathbf{L}_M \mathbf{S}_M \mathbf{R}_2^T} \\ \mathbf{L}_2 \mathbf{S}_M \mathbf{R}_M^T & -\mathbf{L}_2 \mathbf{C}_M \mathbf{R}_2^T \end{bmatrix} \begin{matrix} r_M \\ c_M \end{matrix}, \qquad (5)$$

where $\mathbf{L}_M \rightarrow \mathbf{L}_1$ and $\mathbf{R}_M \rightarrow \mathbf{R}_1$ are the left and right singular vectors of normalised matrix $\mathbf{M}$ partitioned as column matrix upon performing singular value decomposition

$$\mathbf{M} = \mathbf{L}_M \mathbf{\Sigma}_M \mathbf{R}_M^T, \qquad (6)$$

with singular values

$$\mathbf{\Sigma}_M = \text{diag}(\sigma_1, \sigma_2, \cdots, \sigma_{\min(r_M, c_M)}), \quad \text{where} \quad \sigma_1 = 1 \geq \sigma_2, \cdots \geq 0, \qquad (7)$$

from which we construct the cosine and sine matrices,

$$\begin{aligned} \mathbf{C} &\leftarrow \text{diag}(\cos\theta_1, \cos\theta_2, \cdots), \\ \mathbf{S} &\leftarrow \text{diag}(\sin\theta_1, \sin\theta_2, \cdots), \\ \theta_i &= \cos^{-1}(\sigma_i), \quad \text{with} \quad i = 1, 2, \ldots, \min(r_M, c_M). \end{aligned} \qquad (8)$$

Thus, satisfying the $\mathbf{C}^2 + \mathbf{S}^2 = \mathbf{1}$ condition posed on $\mathbf{C}$ and $\mathbf{S}$. By direct calculation, it can be deduced that the two unitary matrices $\mathbf{L}_2$ and $\mathbf{R}_2$ are free variables, as one can choose any unitary matrix to satisfy the CSD theorem. Noting that the variable substitutions done in (5)-(8) have taken advantage of the filler matrices $\mathbf{0}_C$ and $\mathbf{1}_S$ to get the most compact unitary matrix $\mathbf{U}$ with the size of $N = r_M + c_M$. The numerical algorithm of CSD completion is given in Supplementary material S1.

Although we specifically consider real-valued matrix $\mathbf{M}$ due to the elementary nature of operation with a real number, the CSD method can also be directly used to complete a non-unitary complex-valued matrix by replacing the transpose operator with the transpose-conjugate operator. It is important to note that complex-valued matrix multiplication can be represented as operations involving the multiplication and addition of real matrices, indicating the extendibility of the methodology to address complex valued matrix (Golub & Van Loan, 2013).

The next step in design framework proposed in the present work is the successive nulling algorithm that decomposes the unitary matrix obtained from the CSD completion into a multiplication of a set of transfer matrices based on a practical building block. There are variations of such a decomposition algorithm, leading on a variety of physical photonic circuit architecture, such as triangular (Reck et al., 1994) and rectangular configuration (Clements et al., 2016). Without loss of generality, here, we use the rectangular configuration to demonstrate our proposed overall design framework; moreover, it provides a symmetrical architecture and a lower loss compared to the triangular system.

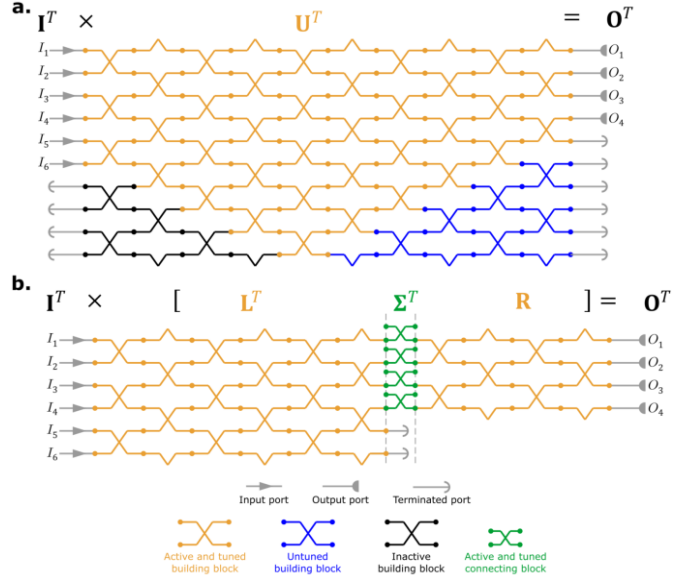

**Figure 2**: Schematic illustration of implementation of arbitrary linear operation of six inputs, $I$, to 4 outputs, $O$, using rectangular photonic circuit architecture. (**a**) Using the proposed design framework of combined CSD and successive nulling algorithm. (**b**) Using direct SVD and successive nulling algorithm.

Figure 2(a) illustrates the rectangular photonic circuit architecture, that implements the operation $\mathbf{O} = \mathbf{UI}$, where $\mathbf{O}$ and $\mathbf{I}$ denote the output and input vectors, respectively. Specifically, it shows the composition of building blocks for implementing an illustrative example of 4-by-6 target matrix $\mathbf{M}$, which upon CSD completion yields a 10-by-10 unitary matrix $\mathbf{U}$. As defined in (5), the matrix $\mathbf{M}$ has been partitioned as the top-left submatrix $\mathbf{U}_{11}$ to operate on the first $c_M$ inputs to the first $r_M$ outputs. Thus, the other ports are not used and should be terminated with matched (impedance) terminations to avoid back reflections. In this example, as shown in Fig. 2(a), only the first 6 input ports and 4 output ports are used. Furthermore, by visual inspection, it can be deduced that because the decomposed rectangular architecture is a unidirectional circuit, not all building blocks need to be tuned. The advantage and trade-offs of implementing the unitary matrix from the CSD completion method, compared to the standard singular value decomposition (SVD) approach (Miller, 2013), in Fig. 2(b), will be further discussed in Section 3.1.

To proceed with the successive nulling algorithm, the transfer function of the unit building block needs to be defined. In the present work, a simple building block consisting of a directional coupler (DC) equipped with a S-bend waveguide is used. Figure 3(a) schematically shows the DC building block with variable length $l_c$ used with an S-bend waveguide located at the top-left arm of the coupler. The length $\mathcal{L}$ of the S bend is designed to realise the required phase shift $\phi$. In the present work, this is done by fixing the horizontal displacement $d_x$ and varying the vertical displacement $d_y$.

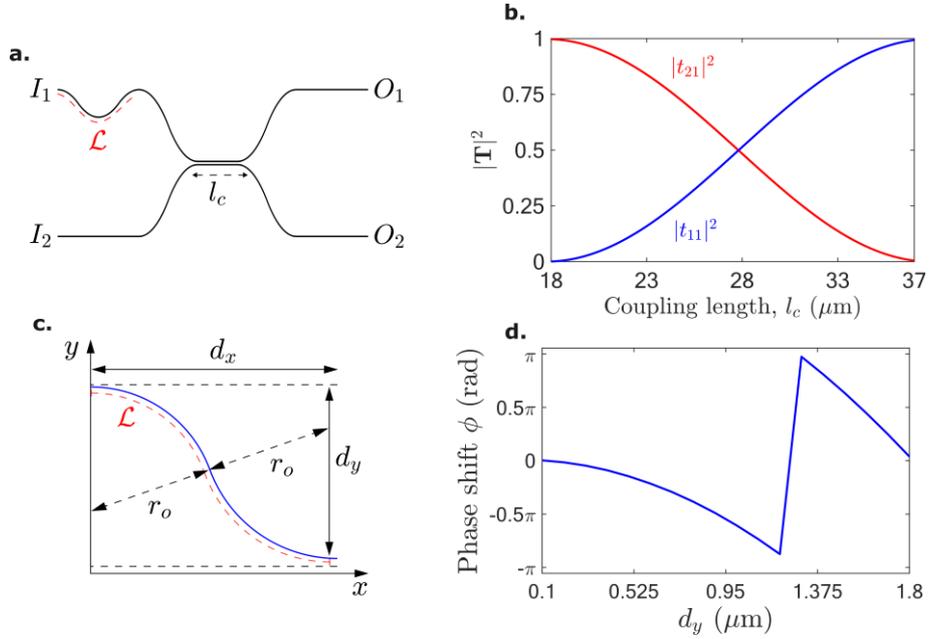

**Figure 3**: (**a**) Building block is comprised of a directional coupler and a S-bend waveguide. (**b**) $t_{21}$ and $t_{11}$ parameter of the DC block as a function of coupler waveguide length $l_c$. (**c**) Detail design of the S-bend waveguide that is defined by vertical displacement $d_y$ and horizontal displacement $d_x$. (**d**) Phase shift $\phi$ as a function of vertical displacement, $d_y$ and fixed horizontal displacement $d_x = 6.585\ \mu m$.

Such a building block provides the ability to independently control both the amplitude and phase-shift and can be modelled using a transfer matrix **T** that relates the inputs and outputs of the block as (Capmany & Pérez, 2020),

$$\begin{bmatrix} O_1 \\ O_2 \end{bmatrix} = \mathbf{T} \begin{bmatrix} I_1 \\ I_2 \end{bmatrix}, \quad \text{with } \mathbf{T} = \begin{bmatrix} e^{j\phi} \cos\theta & j\sin\theta \\ je^{j\phi} \sin\theta & \cos\theta \end{bmatrix}, \quad (9)$$

where $\phi$ is external phase-shift, $\phi \in [0, 2\pi]$ and $\theta = \kappa l_c$ is the DC phase depending on the coupling strength $\kappa$ and the DC length $l_c$, with the bar-transmission coefficient given by $\cos\theta$ and the coupling constant $K = \sin^2\theta$. As described in (Reck et al., 1994), the transfer matrix (9) can be represented as

$$\mathbf{T}_{mn}(\theta, \phi) = \begin{bmatrix} 1 & 0 & \cdots & & \cdots & & \cdots & 0 \\ 0 & 1 & & & & & & \vdots \\ \vdots & & \ddots & & & & & \vdots \\ \vdots & & & e^{j\phi}\cos\theta & & j\sin\theta & & \vdots \\ & & & & \ddots & & & \\ \vdots & & & je^{j\phi}\sin\theta & & \cos\theta & & \vdots \\ & & & & & & \ddots & \\ 0 & \cdots & & \cdots & & \cdots & 0 & 1 \end{bmatrix} \begin{matrix} 1 \\ 2 \\ \vdots \\ m \\ \vdots \\ n \\ \vdots \\ N \end{matrix} \quad (10)$$

to describe operations between the input ports $m$ and $n$ with the output ports $n$ and $m$, respectively. Thus, by using the transfer matrix in (10), any unitary matrix can be diagonalised a (Capmany & Pérez, 2020) (Perez et al., 2017),

$$\left(\prod_{(m,n)\in S_L} \mathbf{T}_{mn}(\theta, \phi)\right) \mathbf{U} \left(\prod_{(m,n)\in S_R} \mathbf{T}_{mn}^{-1}(\theta, \phi)\right) = \mathbf{D}. \quad (11)$$

where $\mathbf{D} = \text{diag}(e^{j\phi_{o1}}, e^{j\phi_{o2}}, \cdots, e^{j\phi_{oN}})$ is a complex-valued diagonal matrix with modulus of one and the indexing $(m,n)$ in $S_L$ and $S_R$ is depending on the order of nulling process, (Wu et al., 2024) (Clements et al., 2016). Numerically, the diagonalisation in (11) is computed by performing a successive nulling of matrix $\mathbf{U}$ by finding the appropriate values of $\theta, \phi$ for each of the building blocks' transfer matrix $\mathbf{T}_{mn}$, as shown in Algorithm 2 in the Supplementary material. This nulling process provides the $\theta, \phi$ parameters required to design each of the $N(N-1)/2$ building blocks.

Using the DC model (9), the physical design parameters of the building block can be obtained, that is the S-bend optical path length $\mathcal{L}$ from $\theta$ and directional coupler length $l_c$ from $\phi$. In this paper the numerical realisation of the photonic circuit is on silicon-on-insulator platform using a single-mode strip buried waveguide operating at 1550 nm. The refractive index of the waveguide core is $n_{\text{silicon}} = 3.478$, the refractive index of the cladding is $n_{\text{silica}} = 1.444$ and the waveguide core width and height are 450 nm.

The transmission coefficients of the DC block having a separation gap of 180 nm are simulated using the full-wave FEM method and shown in Fig. 3(b) as a function of the coupler length $l_c$. Figure 3(c) confirms that the choice of coupler length can fully control the amplitudes of the transmitted signals at outputs $O_1$ and $O_2$ from zero to unity by varying the DC length $l_c$ from 18 µm to 37 µm. The external phase shift $\phi$ is used to design the S-bend using

$$\phi = \beta_{\text{eff}} \mathcal{L} = \beta_{\text{eff}} r_0 \cos^{-1}\left(1 - \frac{d_y}{2r_0(d_y)}\right), \tag{12}$$

where $\beta_{\text{eff}} = 2\pi n_{\text{eff}}/\lambda_0$ is the effective propagation constant, bend radius $r_0$ is obtained using $r_0(d_y) = (d_x^2 + d_y^2)/(4d_y)$. Figure 3(d) shows the range of $\phi$ obtained by varying $d_y$ parameter for $d_x = 6.585\ \mu m$ confirming the ability to achieve a full range of phase shift ($0 < \phi < 2\pi$).

## 3. Results and Discussion

### 3.1 Discussion on the CSD Unitary Completion Approach

Before demonstrating the application of the design framework in Figure 1, it is worth discussing the advantage and trade-offs of the CSD unitary completion approach proposed in this work, compared to the direct SVD approach (Miller, 2013). Specifically on the size of the matrix **U** and its implication for the number of required building blocks.

As detailed in Section 2, for a given matrix $\mathbf{M} \in \mathbb{R}^{r_M \times c_M}$, unitary completion using CSD decomposition will yield to a higher-dimensional square matrix $\mathbf{U} \in \mathbb{R}^{N \times N}$, where $N = r_M + c_M$. As such upon application of successive nulling decomposition with rectangular architecture, as detailed in Section 2, this results in a circuit with $N(N-1)/2$ number of building blocks. In contrast, for the direct application of SVD decomposition on the given matrix $\mathbf{M} \in \mathbb{R}^{r_M \times c_M}$, the decomposition $\mathbf{M} = \mathbf{L}_M \mathbf{\Sigma}_M \mathbf{R}_M^T$ yields two unitary matrices $\mathbf{L}_M \in \mathbb{R}^{r_M \times r_M}$ and $\mathbf{R}_M \in \mathbb{R}^{c_M \times c_M}$. Applying the same rectangular architecture decomposition procedure, on both $\mathbf{L}_M$ and $\mathbf{R}_M$ matrices, results in a circuit with $[r_M(r_M - 1) + c_M(c_M - 1)]/2$ number of building blocks.

Comparing the number of building blocks from both approaches, it can be determined that the SVD approach results in a circuit with fewer building blocks. A comparison of circuit designs for a 6-by-4 matrix **M** implemented using both CSD completion and SVD decomposition is shown in Fig. 2. However, it is important to note that in the circuit from the CSD completion, not all input and output ports are used, leaving a subset of these building blocks inactive. In contrast, the SVD approach results in a circuit with all active building blocks. The key advantage of the CSD completion compared to the SVD approach is that the CSD procedure is a topology-conserving, it allows the use of a single system to implement the whole matrix **U** while the SVD approach due to its specific decomposition results in the need of two systems connected with specialised connecting blocks implementing the diagonal singular values matrix $\mathbf{\Sigma}_M$ as illustrated in Fig. 2(b).

### 3.2 Example Implementation of the Design Framework

In this section, the design framework, i.e. the CSD unitary completion and the successive nulling method for a rectangular photonic circuit architecture, is demonstrated on an exemplary 2-by-2 non-unitary matrix **M**,

$$\mathbf{M} = \begin{bmatrix} 0.920 & 0.087 \\ 0.123 & 0.860 \end{bmatrix}. \tag{13}$$

Note that the matrix **M** has been chosen to have the largest singular value of one. For matrices with singular values larger than one, pre-normalisation is required as described in Section 2. By applying the CSD completion in (5), (Algorithm 1), the matrix **U** is obtained as

$$\mathbf{U} = \left[\begin{array}{cc|cc} 0.920 & 0.087 & 0.311 & -0.219 \\ 0.123 & 0.860 & -0.404 & 0.285 \\ \hline 0.069 & -0.093 & -0.684 & -0.719 \\ -0.364 & 0.493 & 0.520 & -0.594 \end{array}\right], \tag{14}$$

which is achieved using the following decompositions,

$$\mathbf{L}_M = \begin{bmatrix} -0.792 & -0.610 \\ -0.610 & 0.792 \end{bmatrix}, \quad \mathbf{R}_M^T = \begin{bmatrix} -0.804 & -0.594 \\ -0.594 & 0.804 \end{bmatrix},$$
$$\mathbf{C}_M = \begin{bmatrix} 1 & 0 \\ 0 & 0.781 \end{bmatrix}, \quad \mathbf{S}_M = \begin{bmatrix} 0 & 0 \\ 0 & 0.624 \end{bmatrix}, \tag{15}$$

and $\mathbf{L}_2, \mathbf{R}_2$ are taken as random unitary matrices as

$$\mathbf{L}_2 = \begin{bmatrix} 1.883 & -0.185 \\ 0.355 & 0.982 \end{bmatrix}, \quad \mathbf{R}_2^T = \begin{bmatrix} 0.300 & 0.426 \\ -0.817 & 0.576 \end{bmatrix}. \tag{16}$$

The subsequent application of the successive nulling algorithm (Algorithm 2) on the unitary matrix $\mathbf{U}$ results in the coupling constants $K = \sin^2 \theta$, external phase shifts $\phi$, and phase shifts at the output ports $\phi_o$. The schematic of the rectangular architecture of building blocks with parameters, namely phase shifts $\phi$ and coupling constants $K$ generated by the nulling process is shown in Fig. 4(a).

Figure 4(b) shows a schematic of a 4-by-4 multiport photonic circuit based on the building blocks designed based on matrix $\mathbf{U}$ in (14) where phase shifts are realised using a S-bends and curved waveguide sections and coupling constants are realised using DCs. Full-vectorial FEM simulations were used to optimise each of the DC directional couplers ($DC_{1,\ldots,6}$) by adjusting the coupling length $l_c$ and arc lengths $\mathcal{L}$ to achieve the required amplitude and external phase shift, respectively. Additionally, connecting waveguides ($WG_{1,\ldots,4}$) were designed with specific arc lengths to ensure effective zero phase shift. Finally, the length of the terminal waveguide at each output port was designed to correspond with $\phi_{o_{1,\ldots,4}}$.

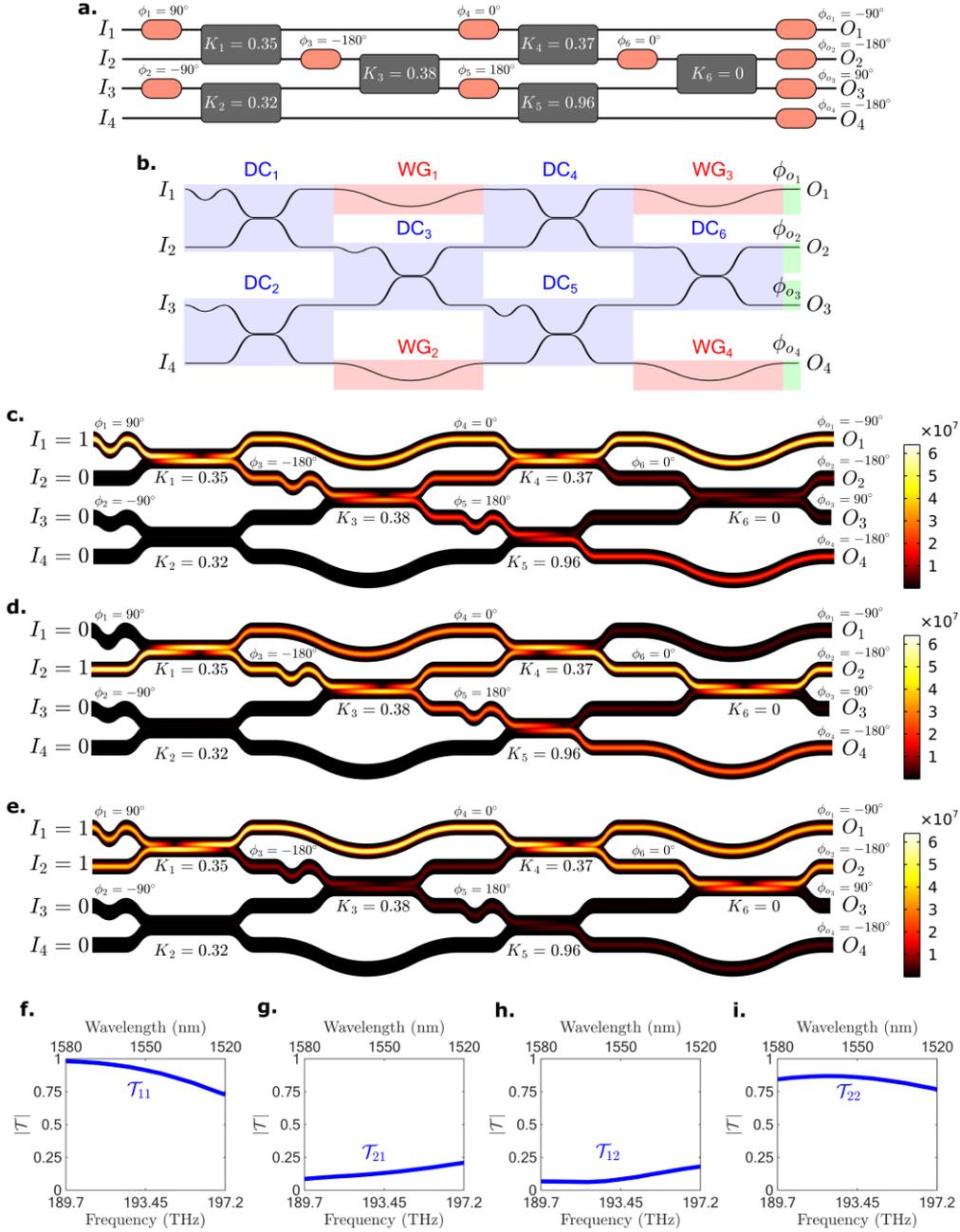

**Figure 4**: (**a**) Schematic of the practical realisation of the 4-by-4 unitary matrix. Coupling constants $K$ and external phase shift $\phi$ for each building block are calculated as in Section 2 (**b**) Schematic of the 4-by-4 photonic circuit where phase shifts are realised using a S-bend and curved waveguide sections and coupling constant is realised using a DC. (**c-e**) Full-wave 3D simulation of the designed 4-by-4 photonic circuit when (c) $I_1$ is on, (d) $I_2$ is on, and (e) $I_1$ with $I_2$ are on. (**f-i**) The transmission parameters $\mathcal{T}_{i,j}$ are entries of $\mathcal{T}$ as a function of frequency (wavelength), where $i,j$ are the output and input ports.

Figures 4(c-e) show the intensity of the optical signal as it propagates through the circuit simulated by FEM for different excitation cases, namely when (c) $I_1$ is on, (d) $I_2$ is on, and (e) $I_1$ and $I_2$ are on. This demonstrates the coherent application of the circuit in transforming the input signal, operating at 1550 nm, described by,

$$\mathcal{T}_{\text{FEM}} = \begin{bmatrix} 0.92\angle-4.72° & 0.08\angle-25.63° & 0.31\angle-5.60° & 0.22\angle174.16° \\ 0.13\angle-20.30° & 0.86\angle-5.01° & 0.35\angle171.50° & 0.33\angle-7.17° \\ 0.08\angle-8.47° & 0.03\angle174.85° & 0.70\angle175.06° & 0.70\angle177.14° \\ 0.36\angle179.51° & 0.49\angle0.34° & 0.52\angle-5.71° & 0.59\angle177.20° \end{bmatrix}. \quad (17)$$

Note that $\mathcal{T}_{\text{FEM}}$ is computed by FEM as a complete circuit and describes the input-output operation, i.e. $\mathbf{O} = \mathcal{T}\mathbf{I}$. Furthermore, the simulated $\mathcal{T}_{\text{FEM}}$ features additional phase shifts, which arise from the fact that the waveguide bending affects the propagation constant and hence this introduces additional phase shifts. This bending effect is captured by the full-wave simulation but not by the transfer matrix model in (10). As a result, the largest eigenvalue is 0.92 in the case of (17). It is noted that although the additional phase shifts can be minimised, for example, by optimising the radial curvature of the S-bend, this will consequently increase the footprint. Nonetheless, it accurately performs the operation expected of the matrix $\mathbf{U}$, with normalised mean squared error NSE of 0.028. NSE is calculated by NSE = $\|\mathcal{T} - \mathbf{U}\|^2/N$, where $\|\cdot\|$ is the L-2 norm.

Figures 4(f-i) show the transmission $\mathcal{T}$ of the 4-by-4 circuit in a narrowband frequency region between 189.7 THz to 197.2 THz, representing a system's spectral response of about 4% bandwidth around the designated operational frequency of 193.4 THz (wavelength of 1550 nm). Special attention is given to the top-left sub-block of $\mathcal{T}_{i,j}$, where $(i,j) \in \{1,2\}$, which implements the target matrix $\mathbf{M}$. These transmission $\mathcal{T}$ are dispersive with variation of the largest variation observed for $\Delta|\mathcal{T}_{1,1}|$ of 27.2% over 4% bandwidth. For completeness, all $\mathcal{T}$-parameter spectra are in Fig. S1 in the Supplementary material.

### 3.3 Analysis of Scalability

Due to limited computational resources, a full-wave 3D FEM simulation is only performed and validated for the 4-by-4 photonic circuit. In this section, we investigate the scalability of the design framework for the larger photonic circuits. To model a larger circuit, a combination of full-wave FEM simulation and circuit design is used whereby each directional coupler and waveguide section is individually optimised for its coupling length $l_c$ and arc lengths $\mathcal{L}$ to satisfy the parameters computed from the nulling process. The S-parameters of these sections are then imported into a circuit modelling tool (Keysight, 2017) to model the entire photonic circuit. As each individual building block is designed separately, it can be done in a parallel manner.

The accuracy of this approach is now analysed by comparing the $\mathcal{T}$-parameters of 4-by-4 unitary matrix computed using a full-wave FEM simulation and using a combined FEM-circuit approach. Using the same matrix **M** as in Section 3.2, the $\mathcal{T}$-parameter computed by the combined FEM-circuit approach is

$$\mathcal{T}_{\text{FEM-circuit}} = \begin{bmatrix} 0.92\angle-4.90° & 0.08\angle-25.85° & 0.31\angle-5.75° & 0.22\angle173.98° \\ 0.13\angle-20.50° & 0.86\angle-5.18° & 0.35\angle171.37° & 0.33\angle-7.34° \\ 0.08\angle-8.67° & 0.03\angle174.68° & 0.70\angle174.86° & 0.70\angle176.91° \\ 0.36\angle179.28° & 0.49\angle0.12° & 0.52\angle-5.90° & 0.59\angle176.96° \end{bmatrix}. \quad (18)$$

Comparing (17) and (18), it can be seen that both approaches agree and thus provide an accurate means to model the circuit, with the NSE for circuit model being 0.027. The $\mathcal{T}_{\text{FEM-circuit}}$ also features a similar amount of additional phase shifts, reaffirming that these phase shifts are due to perturbations occurring inside each of the building blocks and accumulating as the optical signal propagates through the circuit.

The FEM-circuit approach is applied to model rectangular photonic circuits arising from various sizes of matrix **U**, namely, $N = 4, 5, 6,$ and $7$. To ensure replicability, for each matrix size $N$, four independent implementations of randomly generated matrices **M** are realised and modelled using the FEM-circuit approach. For $N = 10$ and $N = 13$, only a single implementation was done due to large number of building blocks needed to be optimised, i.e. 45 and 78 building blocks respectively, and our limited computational resources.

Figure 5(a) shows the NSEs of the $\mathcal{T}$-matrix computed using FEM-circuit approach for different sizes of the unitary matrix **U**. The NSE increases for larger $N$, this trend follows and reaffirms the rationale given in the previous sections, that the error is a result of accumulation of loss and phase addition due to waveguide bendings over the propagation of the optical signal and thus the deeper (number of layers) of the circuit the larger the error it accumulates. In Fig. 5(a), the hollow circle line across the figure represents means NSE for the sizes where multiple results were available. The error bars denote the standard deviation of these results, providing an indication of the variability of the NSE across different implementations.

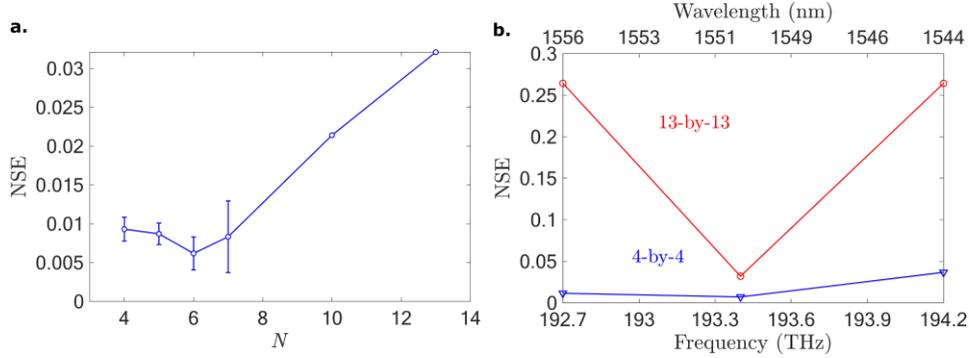

**Figure 5**: (**a**) NSE for different size $N$ of multiport interferometers. The hollow circle line indicates the mean of NSE values while the error bar indicates the standard deviation of NSE values except for $N = 10$ and $N = 13$ where only a single implementation was implemented. (**b**) Frequency response for photonic circuit. The triangular curve is the performance for 4-by-4 multiport interferometer. The hollow circle curve is the performance for 13-by-13 photonic circuit.

Figure 5(b) investigates the robustness of the 4-by-4 and 13-by-13 circuits to small frequency detuning by comparing the accuracy of the operations performed at three frequencies: 192.7 THz, the designed operational frequency of 193.4 THz and 193.4±0.75 THz. While the 4-by-4 circuit maintains a low NSE across these frequencies, indicating robust performance, the 13-by-13 circuit performs with low error (NSE = 0.0321) only at the design frequency 193.4 THz. It shows that the 13-by-13 circuit is sensitive to frequency detuning, i.e., NSE > 0.25 over 0.75 THz detuning. This sensitive behaviour can be attributed to the increased number of dispersive building blocks. As each building block is naturally dispersive, the increase in the number of successive operations using these building blocks amplifies the dispersive effect. For completeness, the full spectral response of $\mathcal{T}$-parameter of the 13-by13 circuit is presented in Fig. S2 in the Supplementary material.

## 4. Conclusion

In summary, we demonstrated a design framework for practically realising any non-unitary matrix multiplication operation using a photonic processor circuit based on directional coupler building blocks. This new design framework utilises CSD theorem to recover the unitarity of a given non-unitary matrix and thus allowing the use of successive nulling decomposition to map a physical realisation. We adapted the rectangular decomposition to design the required physical coupling constants and external phase shifts of the directional coupler building blocks. Both full-wave FEM simulation and the hybrid FEM-circuit method are validated and used to simulate the practical photonic circuit system. The scalability and robustness of the design framework were numerically investigated. Although our study reveals limitations in the scalability of the decomposition approach, it shows that accurate operations are achievable for large circuits at the designed operational frequency, provided that appropriate optimisation of each building block is undertaken in order to minimise overall accumulated error. The study suggests the need for building blocks with reduced propagation constant perturbations while providing the required wave amplitude and phase control

**Acknowledgement**

We thank William Clement for sharing his code and for the fruitful collaboration. H.T. acknowledges the support of Higher Committee for Education Development (HCED) of the Republic of Iraq for the PhD scholarship.